\documentstyle[preprint,aps]{revtex} 
\begin{document} 
\tightenlines
\draft 
\title{Schwinger Model Green functions with topological effects} 
\author{Tomasz Rado\.zycki\thanks{Elecronic mail:
 torado@fuw.edu.pl}} 
\address{Physics Department, Warsaw University, ul. Ho\.za 69, 
00-681 Warsaw, Poland} 
\maketitle 
\begin{abstract} 
The fermion propagator and the 4-fermion Green function in the massless 
QED$_2$ are explicitly found with topological effects taken into account. 
The corrections due to instanton sectors $k=\pm 1$, contributing to the 
propagator, are shown to be just the homogenous terms admitted by the 
Dyson-Schwinger equation for $S$. In the case of the 4-fermion function also 
sectors $k=\pm 2$ are included into consideration. The quark condensates 
are then calculated and are shown to satisfy cluster property. The 
$\theta$-dependence exhibited by the Green functions corresponds to
and may be removed by performing certain chiral gauge transformation.
\end{abstract} 
\pacs{11.10.-z;11.30.Rd;12.90.+b} 

\section{Introduction} 
\label{sec:intro} 

Many phenomena in present-day theory of elementary particles can be 
only described in the nonperturbative language. In the first place (though 
not exclusively) it refers to the theory of strong interactions --- quantum 
chromodynamics --- where the only existence of mesons and baryons requires 
allowing for such effects as quark and gluon confinement. However QCD, 
based on colour $SU(3)$ symmetry group, introduces a variety of 
coupled fields, which makes the perturbative calculation very complicated, 
and nonperturbative one extremely perplexed.

Fortunately we have at our disposal a couple of examples of model 
theories, that, thanks to their mathematical simplicity, allow for 
investigating certain nontrivial and nonperturbative aspects of more 
complicated phenomena. Among most fertile ones one should mention the 
so called Schwinger Model --- the electrodynamics of massless fermions 
that, in this work, will be called `quarks', in two space-time dimensions. 
This model was originally proposed~\cite{jsch} as an example of a field 
theory in which the existence of the local gauge symmetry does not entail 
the zero mass gauge boson --- a fact which should not be pushed aside in 
electroweak interactions theory. It turned out soon, however, that it has 
many other nontrivial properties, which can be interesting from the point 
of view of both, strong and weak interactions. Above all one should 
mention here the confinement of fundamental fermions and the axial 
anomaly.

The similarity between quantum chromodynamics and the Schwinger 
Model appears also (and this will be particularly essential for this work) in 
the presence of topological effects: the existence of instanton sectors 
leading to the emerging of $\theta$-vacuum and nonzero `quark' 
condensate.

It is a known fact~\cite{thoo1} that if the theory contains massless 
fermions (at least one) the amplitude of the tunnelling transition between 
different topological vacua is diminished to zero. Mathematically it is 
expressed through zero value of the Euclidean Dirac operator determinant: 
$\det [i\not\! \partial - e\not\!\! A]$, because of the appearance of
the zero 
eigenvalues and eigenvectors when $A$ bears the instanton 
index~\cite{as,thoo2,kis,ans,huang}. The functional integral over 
fermionic degrees of freedom, corresponding to the transition amplitude in 
question, is then zero and the tunnelling phenomenon disappears. On the 
other hand, however, the notion of the $\theta$-vacuum does not lose its 
sense since topological vacua do not exhibit the so called cluster 
property~\cite{cdg}. The reason is that, despite the appearance of the 
anomaly, a conserved chiral charge still can be defined, the topological 
vacua being its eigenstates with different eigenvalues. The matrix elements 
$<m|Q(x)|n>$ of certain local operator $Q(x)$ between such vacua 
($|n>$ and $|m>$) are then  nonvanishing if the operator $Q$ 
changes the chirality (the Hamiltonian does not belong to that type of 
operators and hence the above mentioned lack of tunnelling). If one of the 
topological vacua were taken as the true vacuum, for such a type of 
operators $Q$ one would obtain automatically $<Q(x)>_{vac} = 0$. A 
product of such operators $P(x,y)=Q_1(x)Q_2(y)$ can, however, have 
the nonzero vacuum expectation value, for instance if it is  constructed in 
a chiral invariant way (e.g. if $Q_1$ changes the chirality by $k$
and $Q_2$ by $-k$). Now the requirement of clusterization in the
limit $|x-y|\rightarrow \infty$ leads to 
contradiction~\cite{raja}
$$
0\neq <P>_{vac}=<Q_1>_{vac}\cdot <Q_2>_{vac} = 0\; .
$$
If we then have to do 
with vacuum expectation values for chirality nonconserving operators, 
we have to include in the calculation different topological 
vacua. It is just that category that fermion, and fermion-boson Green 
functions belong to.
Because of the $2 : 1$ correspondence between the chiral charge and the 
topological index of a vacuum, the objects bilinear in fields $\Psi$ 
(propagators) require the inclusion of the instanton sectors $0,\pm 1$, and 
4-fermion functions --- sectors $0,\pm 1,\pm 2$.

All these facts are, of course, well known and applied successfully to the 
calculations of the quark condensate $<\overline{\Psi}(x)\Psi(x)>_{vac}$ 
in the Schwinger Model, as well as to the verification of the revival of 
the cluster property for $<\overline{\Psi}(x)\Psi(x) 
\overline{\Psi}(y)\Psi(y)>_{vac}$~\cite{cadam1,smil,gattr}.

The topological methods, however, have not been till now applied to 
the calculation of the full 2- and 4-point Green functions and not only 
condensates. This computation will constitute the main goal in the
present work. The paper is organized in the following way. In
Section~\ref{sec:2p} we concentrate on 
the quark propagator. From the considerations of the Dyson-Schwinger 
equations one knows~\cite{stam,thzh} that, beside the famous Schwinger 
solution~\cite{jsch} satisfying $\{S,\gamma^5\}=0$ i.e. $S\sim 
\gamma^{\mu}(...)_{\mu}$, they admit also other terms, let us call them $S'$, 
for which $[S',\gamma^5]=0$. That means that $S'\sim \gamma^5(...) + 
\openone (...)$. The authors raised then the question concerning the 
interpretation of such terms. Extending slightly the analysis of~\cite{thzh} 
we show, in Section~\ref{subsec:general}, which the most general form of the
quark propagator, resulting from the Dyson-Schwinger equations, is.
Next, in Section~\ref{sec:2pinst} using the methods 
of~\cite{cadam1}, applied there to find the value of condensate, we calculate 
the full propagator and show that the additional terms $S'$ are just those 
emerging from nonzero instanton sectors. In section~\ref{sec:4p} we perform 
the similar calculation for the 4-fermion function. The analysis is here 
more complicated since it now involves the $0,\pm 1,\pm 2$ instanton 
sectors and also because of the tensor structure of $G$, which bears
now the four spinor indices. In the final section we verify the
cluster property and 
have a look on the $\theta$ dependence of the full Green functions. As one 
knows, in massless theory the parameter $\theta$ may be removed by the 
convenient chiral gauge transformation. This conclusion will find its full 
confirmation in our expressions for the Green functions.

\section{Instanton contributions to the quark propagator}
\label{sec:2p}

In this section we would like to concentrate on the fermion propagator. 
Being defined as the vacuum expectation value of the product of $\Psi$ 
and $\overline{\Psi}$ it should acquire additional terms, beside that found 
already by Schwinger, originating from instanton sectors $\pm 1$. 

We start with summarizing briefly the conventions used in this work. 
The two-dimensional Lagrangian density of the Schwinger Model with
the gauge fixing term has the following form
\begin{equation} 
{\cal L}(x)=\overline{\Psi}(x)\left(i\gamma^{\mu}\partial_{\mu} -
e\gamma^{\mu}A_{\mu}(x)\right)\Psi (x)-
\frac{1}{4}F^{\mu\nu}(x)F_{\mu\nu}(x)-
\frac{\lambda}{2}\left(\partial_{\mu}A^{\mu}(x)\right)^2\; , 
\label{lagr} 
\end{equation} 
where for Dirac matrices $\gamma^{\mu}$ we assume the representation
in which all gamma's are real
\begin{equation} 
\gamma^0=\left(\begin{array}{lr}0 & \hspace*{2ex}1 \\ 1 & 0 
\end{array}\right)\; , \;\;\;\;\; 
\gamma^1=\left(\begin{array}{lr} 0 & -1 \\ 1 & 0 
\end{array}\right)\; , \;\;\;\;\; 
\gamma^5=\gamma^0\gamma^1=\left( 
\begin{array}{lr} 1 & 0 \\ 0 & -1 \end{array}\right) \; ,
\label{gam}
\end{equation}
The metric tensor $g^{\mu\nu}$ and the totally antisymmetric symbol 
$\varepsilon^{\mu\nu}$ are defined as follows
$$
g^{00}=-g^{11}=1\; ,\;\;\;\;\;  \varepsilon^{01}=-\varepsilon^{10}=1\; 
,\;\;\;\;\; 
\varepsilon^{00}=\varepsilon^{11}=0\; . 
$$

\subsection{General considerations}
\label{subsec:general}
Having specified the Lagrangian we can derive, in the standard 
way, the Dyson-Schwinger equations~\cite{ds1,ds2} for the 
propagators for the two basic fields in the theory. It is well known that the 
result for the gluon Green function is simply that it acquires a mass equal 
to $\sqrt{e^2/\pi}$. We recall that in two dimensions $e$ is a
dimensionful constant. No other contributions from the nonzero instanton 
sectors come into play since $A^{\mu}$ does not change the chiral
charge, similarly as interactions with photons do not change the
electric charge (which is no more true in nonabelian theories).
We, therefore, take the full, dressed gluon
propagator as already known to be (in momentum space) 
\begin{equation}  
D^{\mu\nu}(k) =  \left(g^{\mu\nu}-  
\frac{k^{\mu}k^{\nu}}{k^2}\right)\frac{1}{e^2/\pi-k^2}- 
\frac{1}{\lambda}\frac{k^{\mu}k^{\nu}}{(k^2)^2}\; ,  
\label{fo}  
\end{equation} 
and concentrate only on the quark sector. The Dyson-Schwinger 
equation, as always, chains up the 2- and 3-point functions which  
recurrence (continued to infinity) usually efficiently precludes us
from solving it
\begin{equation}
\not\! p S(p) = 1 + ie^2\gamma^{\mu}\int\frac{d^2k}{(2\pi )^2}S(p+k)
\Gamma^{\nu}(p+k,p)S(p)D_{\mu\nu}(k)\; .
\label{dyson}
\end{equation}
The prominent and well known advantage of the Schwinger Model is that 
its Lagrangian is invariant under two types of gauge transformations: 
ordinary and chiral ones. This leads to two kinds of Ward 
identities~\cite{stam,deth,tho,trjmn} which relate the projections of 
$\Gamma^{\mu}(p+k,p)$ on $k^{\mu}$ and on 
$\varepsilon^{\mu\nu}k_{\nu}$ with quark propagator. In 
two-dimensional world they are sufficient to reconstruct the full vertex 
$\Gamma^{\mu}$ and decouple the equation from the infinite hierarchy
\begin{equation}
S(p+k)\Gamma^{\nu}(p+k,k)S(p)=\frac{k^{\nu}}{k^2}\left[S(p)-
S(p+k)\right]-
\frac{\varepsilon^{\nu\alpha}k_{\alpha}}{k^2}\left[\gamma^5S(p)+S(p+
k)\gamma^5\right]\; .
\label{vert}
\end{equation}
 If we now adopt the Landau gauge ($\lambda\rightarrow\infty$) we see 
that the longitudinal part of~(\ref{vert}) does not contribute since
$D^{\mu\nu}(k)$ becomes purely transverse and 
equation~(\ref{dyson}) may be given the following closed form
\begin{equation}
\not\! p S(p)=1-ie^2\int\frac{d^2k}{(2\pi)^2}\frac{1}{k^2(e^2/\pi-k^2 -
i\epsilon)}\not\! k\gamma^5\left[\gamma^5S(p)+S(p+k)\gamma^5\right]\; 
,\label{dyso1}
\end{equation}
where the relation $\varepsilon^{\mu\alpha}\gamma_{\mu}k_{\alpha} 
=\not\! k\gamma^5$ has been used. In the co-ordinate space there is a 
known factorization and the equation is simplified to
\begin{equation}
i\not\!\partial_x S(x)=\delta^{(2)}(x)-e^2 \left[\not\!\partial_x
\beta(x)\right] \gamma^5 S(x) \gamma^5
\label{simdys}
\end{equation}
The function $\beta$ is  here defined as follows
\begin{eqnarray}
\beta(x)&&=\int\frac{d^2p}{(2\pi)^2}\left(1-e^{ipx}\right)
\frac{1}{(p^2-e^2/\pi
+i\epsilon)(p^2+i\epsilon)}=\label{beta}\\
&&\nonumber\\ 
&&=\left\{\begin{array}{ll}\frac{i}{2e^2}\left[-
\frac{i\pi}{2}+\gamma_E
+\ln\sqrt{ e^2x^2/4\pi}+ 
\frac{i\pi}{2}H_0^{(1)}(\sqrt{e^2x^2/\pi})\right] & \hspace*{3ex}
x\;\;\;\; {\rm timelike}\\ 
\frac{i}{2e^2}\left[\gamma_E+\ln\sqrt{-e^2x^2/4\pi}+K_0(\sqrt{-
e^2x^2/\pi})
\right] &  \hspace*{3ex}x\;\;\;\; {\rm spacelike}\; ,\end{array}\right.
\nonumber
\end{eqnarray}
and is, in fact, a function of $x^2$ only. Symbol $\gamma_E$ denotes 
the Euler constant and functions
$H_0^{(1)}$ and $K_0$ are Hankel function of the first kind, and Basset
function respectively~\cite{old}. Assume now the most general matrical 
structure that is admitted in two dimensions, in co-ordinate space, for
the fermion propagator $S$
\begin{equation}
S(x)={\cal 
S}_0(x)A(x^2)+B(x^2)+\gamma^5C(x^2)+\gamma^5\not\!xD(x^2)\; .
\label{asump}
\end{equation}
The term proportional to $\not\! x$ has been chosen, for convenience, as 
explicitly containing the free propagator ${\cal S}_0(x)=-
\frac{1}{2\pi}\frac{\not\!\!\;\, x}{x^2-i\varepsilon}$. It is, of
course, only a 
question of redefining the coefficient function $A(x^2)$. In what follows 
we will omit epsilons specifying that we have to do with causal 
propagator. We now insert~(\ref{asump}) into the co-ordinate space 
Dyson-Schwinger 
equation~(\ref{simdys}) and exploit the fact that for all unknown 
functions, as well as for $\beta$ function, we can write $\not\! \partial_x 
F(x^2)=2\not\! x dF/dx^2 = 2\not\! x F'(x^2)$. This allows us to 
deduce the set of four differential equations for four functions to
be found. This set arises if one takes the trace of~(\ref{simdys})
with successive insertion of $\openone $, $\gamma^5$, $\not\! x$ and
$\gamma^5\not\! x$ on both sides. These are simple first order equations
\begin{equation}
A'=-ie^2\beta'A\; ,\;\;\;\; B'=ie^2\beta'B\; ,\;\;\;\;
C'=ie^2\beta'C\; ,\;\;\;\; D'=-\left[\frac{1}{x^2}+ie^2\beta'\right]D\; ,
\label{difeq}
\end{equation}
with the initial condition $A(0)=1$ originating from the cancellation of 
the Dirac delta functions in~(\ref{simdys}). This set of equations may
immediately be solved and we obtain the most general form of the
propagator that is accepted by the Dyson-Schwinger equation
\begin{equation}
S(x)={\cal S}_0(x)e^{-ie^2\beta(x)}+B_0e^{ie^2\beta(x)}+C_0 
\gamma^5 
e^{ie^2\beta(x)}+D_0\frac{\gamma^5\not\! x}{x^2}e^{-ie^2\beta(x)}\; ,
\label{fulls}
\end{equation}
where constants $B_0, C_0, D_0$ remain unknown.

In what was stated above we did not move to far from what had been 
done in~\cite{thzh}, except that we found two additional terms ($C$
and $D$) in the most general structure of $S$. In what follows,
however, we will show that the terms with unknown constants in~(\ref{fulls}),
except for the last one, arise as a result of instanton effects.
The first term is just the well known Schwinger solution.

\subsection{Explicit calculation of the quark propagator}
\label{sec:2pinst}

The instanton contributions to the quark condensate were calculated in a 
systematic way in~\cite{cadam1}. Below we extend this calculation and 
find the expression for the full propagator. We begin with substituting into 
the generating functional, defined as usually as
\begin{equation}
Z[\eta,\overline{\eta},J]=\int{\cal D}\Psi{\cal D}\overline{\Psi}{\cal 
D}Ae^{i\int d^2x \left[{\cal 
L}+\overline{\eta}\Psi+\overline{\Psi}\eta+J^{\mu}A_{\mu}\right]}\; ,
\label{genfunc}
\end{equation}
the following form of the gauge potential
\begin{equation}
A^\mu(x)=A^{(0)\mu}(x)+\varepsilon^{\mu\nu}\partial_{\nu}b(x)\; ,
\label{genpot}
\end{equation}
where $A^{(0)\mu}$ is certain new potential satisfying some special 
conditions at space-time infinity (this point will be discussed
afterwards) and 
$b$ is the external scalar function to be chosen later for our convenience. 
Since~(\ref{genpot}) constitutes a simple shift we can now easily pass 
from the functional integral over $A$ to that over $A^{(0)}$. 
It is known that the coupling term $e\overline{\Psi}\not\!\! 
A^{(0)}\Psi$ may be gauged away if we introduce new fermion fields 
defined by the relations
\begin{eqnarray}
\Psi(x)&=&e^{-ie\not\!\: \partial_x\int d^2z\bigtriangleup(x-
z)\gamma^{\mu} A^{(0)}_{\mu}(z)}\Psi'(x)\stackrel{\rm def}{=}e^{-
i\phi(x,A^{(0)})}\Psi'(x)\; , \label{gauge1}\\
\overline{\Psi}(x)&=&\overline{\Psi}'(x)e^{ie\gamma^{\mu}\not\!\: 
\partial_x\int d^2z\bigtriangleup(x-z) 
A^{(0)}_{\mu}(z)}\stackrel{\rm
def}{=}\overline{\Psi}'(x)e^{i\tilde{\phi}(x,A^{(0)})}\; , 
\label{gauge2}
\end{eqnarray}
with $\bigtriangleup(x-z)$ satisfying: $ \Box_x\bigtriangleup(x-
z)=\delta^{(2)}(x-z)$. The above gauge transformation is an element of 
$U(1)\otimes U_A(1)$ group. While the Lagrangian~(\ref{lagr}) is 
invariant (apart from the gauge fixing term) under that kind of 
transformations, the fermion path integral measure in~(\ref{genfunc}) is 
not~\cite{fuji,ros,bert}. This anomalous behaviour generates a mass term 
for the gauge boson. After this transformation is performed
in~(\ref{genfunc}), the $A^{(0)}$ dependence 
reappears in the source terms through the functions $\phi (x,A^{(0)})$ and 
$\tilde{\phi}(x,A^{(0)})$ defined by equations~(\ref{gauge1},\ref{gauge2}). 
The substitution $A^{(0)}\rightarrow 
\frac{\delta}{i\delta J}$ allows us to perform the remaining gaussian 
integral over $A^{(0)}$~\cite{com1} and we obtain the following
expression for $Z$
\begin{eqnarray}
&&Z[\eta,\overline{\eta},J]=N_b\exp\left\{i\int 
d^2x\left[\frac{1}{2}b\Box^2b+ 
\varepsilon^{\mu\nu}J_{\mu}\partial_{\nu}b\right]\right\}\cdot \int{\cal 
D}\Psi{\cal D}\overline{\Psi}\times\nonumber \\ &&\exp\left\{i\int 
d^2x\left[\overline{\Psi}\left(i\not\! \partial -
e\varepsilon^{\mu\nu}\gamma_{\mu}\partial_{\nu}b\right)\Psi+
\overline{\eta}e^{-i\phi(x,\delta/i\delta J)}\Psi + 
\overline{\Psi}e^{i\tilde{\phi}(x,\delta/i\delta J)} \eta\right]\right\}\times
\label{zb}\\
&&\exp\left\{-\frac{i}{2}\int d^2xd^2y\left[J^{\mu}+(\Box+ 
e^2/\pi)\varepsilon^{\mu\alpha}\partial_{\alpha}b\right] 
\bigtriangleup_{\mu\nu}(x-y,e^2/\pi) \left[J^{\nu}+(\Box+
e^2/\pi)\varepsilon^{\nu\beta}\partial_{\beta}b\right] \right\}\; , \nonumber
\end{eqnarray}
where $N_b$ is certain constant ($b$ stands for {\em boson}).
The massive propagator $\bigtriangleup^{\mu\nu}(x-y,e^2/\pi)$ satisfies 
the equation
\begin{equation}
\left[\left(\Box+\frac{e^2}{\pi}\right)\left(g_{\mu\nu}-
\frac{\partial_{\mu}\partial_{\nu}}{\Box}\right)+\lambda\partial_{\mu}
\partial_{\nu}\right]\bigtriangleup^{\nu\alpha}(x-y,e^2/\pi)=\delta^{(2)}(x-
y)g_{\mu}^{\alpha}\; .
\label{massd}
\end{equation}

The main point is now the evaluation of the fermion path
integral in~(\ref{zb}). Let us denote by $X$ the following expression
\begin{equation}
X=\int{\cal D}\Psi{\cal D}\overline{\Psi}\exp\left\{i\int 
d^2x\left[\overline{\Psi}\left(i\not\! \partial -
e\varepsilon^{\mu\nu}\gamma_{\mu}\partial_{\nu}b\right)\Psi+
\overline{\eta}'\Psi + \overline{\Psi}\eta'\right]\right\}\; ,
\label{x}
\end{equation}
where primes are used to avoid writing explicitly the factors $e^{-i\phi}$ 
multiplying the external sources. This expression is naturally 
proportional to the determinant of the Dirac operator. From~(\ref{x}) it is 
evident, however, that it strongly depends on the choice of hitherto 
undefined function $b$. It turns out that for certain choices of $b$ it 
may even vanish so one could proceed here with caution. 
First of all we temporarily go over to the Euclidean space since  
the properties of the Dirac operator are there mathematically more 
rigorous. The transition to this space is defined by the following 
substitutions: $x^0\rightarrow -ix_2$, $A^0\rightarrow iA_2$, 
$\partial^0\rightarrow i\partial_2$. The Euclidean metric tensor 
$g_{\mu\nu}$ is defined as $\delta_{\mu\nu}$ and for 
$\varepsilon_{\mu\nu}$ we have $\varepsilon_{12}=-\varepsilon_{21} =-
i$. In what follows the same symbols as before will be used for denoting 
the Euclidean counterparts of earlier defined quantities. We hope that it 
will not cause any confusion and the passing back to the Minkowski space 
will explicitly be stated. Before we calculate the quantity~(\ref{x}) we 
have to make some  remarks on the instanton sectors.

Consider the most general form of the field $A$ in the 2-dimensional world
\begin{equation}
A_\mu(x)=\partial_{\mu}a(x)-\varepsilon_{\mu\nu}\partial_{\nu}b(x)\; ,
\label{ab}
\end{equation}
The so called Pontryagin index for the gluon field may be defined as
\begin{equation}
\nu = \frac{ie}{4\pi}\int d^2x\varepsilon_{\alpha\beta}F_{\alpha\beta}= -
\frac{ie}{2\pi}\int d^2x\Box_x b(x)\; .
\label{pontr}
\end{equation}
Due to the Euclidean nature of the space-time the d'Alambert operator 
$\Box_x$ is here, naturally, the same as the Laplace operator. Assume 
now that we take the function $b(x)$ in certain specific form
\begin{equation}
b^{(k)}(x)=\frac{i}{2e}k\ln\left(\frac{x^2+\lambda^2}{\lambda^2}\right)
\; ,
\label{b}
\end{equation}
where $k$ is an integer number. After evaluation of~(\ref{pontr}) we 
immediately obtain $\nu=k$ i.e. the field $A^{(k)}_{\alpha}$ defined as 
$\varepsilon_{\alpha\beta}\partial_{\beta}b^{(k)}$ bears the 
index $\nu=k$. Since the Pontryagin index~(\ref{pontr}) is linear in 
gluonic field the total index of the sum 
\begin{equation}
A^{(0)}_{\alpha}+\varepsilon_{\alpha\beta}\partial_{\beta}b^{(k)}\; ,
\label{afield}
\end{equation}
is equal to $k$, and thanks to the complete freedom while choosing 
$A^{(0)}$ (which is restricted to the sector 0) constitutes the most
general form of the field in the $k$ 
instanton sector: $A^{(k)}_{\alpha}$. $A^{(k)}_{\alpha}$ 
represents a path (in
the sense of Feynman path integal) connecting two distinct and
topologically inequivalent vacua for $t=\pm\infty$. The true vacuum
of the theory, the 
so called $\theta$-vacuum, is now the superposition of these topological 
vacua
\begin{equation}
|\theta>=\sum_{n=-\infty}^{\infty}e^{in\theta}|n>\; ,
\label{vacuum}
\end{equation}
and the generating functional calculated in this new vacuum has the form
\begin{equation}
Z[\eta ,\overline{\eta}, J]=\sum_{k=-
\infty}^{\infty}e^{ik\theta}Z^{(k)}[\eta ,\overline{\eta}, J]\; ,
\label{gener}
\end{equation}
where the summation runs over instanton indices rather than over 
topological vacua ones. In each sector the appropriate $Z^{(k)}$ is 
calculated with the restriction on the values of the vector potential
to those defined by~(\ref{afield}). Now we are ready for considering the 
contributions from separate terms of the sum in~(\ref{gener}) --- the 
different instanton sectors. In compliance with what was said in the 
Introduction about chirality conserving and nonconserving operators and 
their matrix elements taken between different topological vacua, we have 
$Z[0,0,0]=Z^{(0)}[0,0,0]\stackrel{\rm def}{=} N_f\cdot N_b$ and for
the propagator we can immediately write
\begin{equation}
S(x-y)=\sum_{k=-\infty}^{\infty}S^{(k)}(x-y)=-i\frac{1}{Z^{(0)}[0,0,0]} 
\sum_{k=-\infty}^{\infty}e^{ik\theta}\left.\frac{\delta^2 
Z^{(k)}[\eta,\overline{\eta},J]}{\delta\overline{\eta}(x)\delta\eta(y)}
\right|_{J,\eta,\overline{\eta}=0}\; .
\label{prop}
\end{equation}
Similar formulae are valid also for other Green functions and will be 
exploited in the following section while dealing with 4-point functions. 
The sum in~(\ref{prop}) formally extends to $+\infty$ but practically 
depends on the properties of the operator the vacuum expectation value 
of which is being considered. In the case of  a propagator this operator is 
simply a product $\Psi\overline{\Psi}$ and the whole sum reduces to 
three terms corresponding to $k=0$, $k=+1$ and $k=-1$ which means 
that chiral charge of the contributing vacua can, at most, differ by 2. It is 
not difficult to observe that the case $k=0$ corresponds to the well 
known Schwinger solution~\cite{jsch}, since it simply requires putting 
$b^{(k)}=0$, and is represented by the first term of~(\ref{fulls}). The 
nontrivial topological effects come from two other sectors. We 
concentrate below on the sector $k=+1$, the calculation for $k=-1$ being 
analogous.  The applied method is here that of~\cite{cadam1} and 
therefore we do not plunge into details and point out only the main steps. 

For $k=1$ we have to substitute $b=b^{(1)}$ into~(\ref{x}) (or rather 
into its Euclidean version) and perform the $\Psi$ and $\overline{\Psi}$ 
functional integral. By the Atiyah-Singer index 
theorem~\cite{as,thoo2,kis,ans,huang} the massless Dirac operator in the 
external field bearing the nonzero Pontryagin number possesses zero 
modes. For $k=\pm 1$ there is only such mode and, for gamma 
matrices conventions defined by~(\ref{gam}), it has the form (for 
the discussion and construction of the zero modes in the Schwinger 
Model see~\cite{cadam1,maie,jaye})
\begin{eqnarray}
\chi_0(x)&=&\frac{1}{\sqrt{2\pi}}\left(\frac{1}{\lambda^2+x^2} 
\right)^{1/2}\left(\begin{array}{c}0 \\ 1\end{array}\right)\;\;\; {\rm for}
\; k=+1\; ,\label{zerop}\\
\chi_0(x)&=&\frac{1}{\sqrt{2\pi}}\left(\frac{1}{\lambda^2+x^2} 
\right)^{1/2}\left(\begin{array}{c}1 \\ 0\end{array}\right)\;\;\;
{\rm for}\; k=-1\; ,\label{zerom}
\end{eqnarray}

In the 2-instanton sector, which will be dealt with in Section~\ref{sec:4p},  
two zero modes appear
\begin{eqnarray}
\chi_0(x)&&=\frac{1}{\sqrt{2\pi}}\frac{\lambda^{3/2}}{\lambda^2+x^2} 
\left(\begin{array}{c}0 \\ 1\end{array}\right)\; , \;\; 
\chi_1(x)=\frac{1}{\sqrt{2\pi}}(x_1-
ix_2)\frac{\lambda^{1/2}}{\lambda^2+x^2} \left(\begin{array}{c}0 \\ 
1\end{array}\right)\;\;\; {\rm for}\; k=+2\; ,\label{2zerop}\\
\chi_0(x)&&=\frac{1}{\sqrt{2\pi}}\frac{\lambda^{3/2}}{\lambda^2+x^2
}\left(\begin{array}{c}1 \\ 0\end{array}\right) \; ,
\;\; 
\chi_1(x)=\frac{1}{\sqrt{2\pi}}(x_1+ix_2)\frac{\lambda^{1/2}}{\lambda^2+
x^2} \left(\begin{array}{c}1 \\ 0\end{array}\right) \;\;\; {\rm
for}\; k=-2\; 
,\label{2zerom}
\end{eqnarray}

The whole set of eigenmodes will be denoted by $\chi_n$. We have
\begin{equation}
\Psi=\sum_n a_n\chi_n\; ,\;\;\;\; \overline{\Psi}=\sum_n b_n\chi_n^+\; ,
\label{expan}
\end{equation}
where $a_n$ and $b_n$ are the Grassmann coefficients. We now pass 
from the integration over $\Psi$ and $\overline{\Psi}$ to that over $a_n$ 
and $b_n$. The presence of a zero mode means that neither $a_0$ nor 
$b_0$ appear in the bilinear term in the exponent of~(\ref{x}) and only 
source terms in the integration over these two parameters are involved. 
The rules for the Grassmann integration demand, however, that from the 
whole exponent only the linear part contribute. The integration over 
nonzero modes may be done in usual way, the details of which can be 
found in~\cite{cadam1}. In that way we arrive at the following expression 
for the quantity $X$
\begin{eqnarray}
X=&&N_f\lambda\int 
d^2xd^2y\overline{\eta}'(x)\chi_0(x)\chi_0^+(y)\eta'(y)\exp\bigg[-\int
d^2xd^2y\overline{\eta}'(x)e^{-i\phi(x,-\varepsilon_{\mu\nu} 
\partial_{\nu}b^{(1)})}\nonumber \\
&&S_0(x-y)e^{i\tilde{\phi}(y,-\varepsilon_{\mu\nu} 
\partial_{\nu}b^{(1)})}\eta'(y)\bigg]\cdot\exp\left[-\int 
d^2x\frac{e^2}{2\pi}b^{(1)}\Box b^{(1)}\right]\; ,
\label{x1}
\end{eqnarray} 
$S_0$ being the Euclidean free fermion propagator: $i\not\!
\partial_x S_0(x-y) = - \delta^{(2)}(x-y)$. The additional factor
$\lambda$ arises from the change of 
variables~(\ref{expan}) since the dimension of field $\Psi$ in the
Feynman path integral measure is $1/2$ and that of $a_0$ and $b_0$
--- 0. 

Now, when the integration over fermionic degrees of freedom has been 
performed, and the properties of the Euclidean Dirac operator have been 
exploited, it is convenient to come back to Minkowski space, where the 
final formula for the propagator, analogous to~(\ref{fulls}), is to be 
obtained. It should, however, be emphasized that whenever required by 
the mathematical rigour, the corresponding expressions in Euclidean 
space are presumably considered.

The differentiation over external currents $\eta$ and $\overline{\eta}$, as 
required by~(\ref{prop}), leads to a very simple expression if one keeps in 
mind that finally we have to set all the external sources to zero
\begin{equation}
\left.\frac{\delta^2 X}{\delta \overline{\eta}(x)\delta\eta 
(y)}\right|_{\eta,\overline{\eta}=0}=-N_f\lambda e^{-i\phi(x,\delta/i\delta 
J)}\chi_0(x)\chi_0^+(y)e^{i\tilde{\phi}(y,\delta /i\delta J)}e^{i\int 
d^2x\frac{e^2}{2\pi}b^{(1)}\Box b^{(1)}}
\label{difx}
\end{equation}
From the definition~(\ref{zerop}) of the zero mode it is evident that the 
product $\chi_0\chi_0^+$ has the matrical form 
$\frac{1}{2}\left(\openone -\gamma^5\right)$. The functional $\phi$ has 
similar structure (a linear combination of $\openone $ and $\gamma^5$) 
and therefore both quantities commute with each other which leads to the 
considering of the operator
\begin{equation}
\exp\left[-i\phi(x,\delta/i\delta J)+i\tilde{\phi}(y,\delta/i\delta
J) \right]\; ,
\label{oper}
\end{equation}
where functional differentiations act on the $J$ dependent argument of 
the last exponent function in~(\ref{zb}). Let us now introduce the 
(nonconserved) current ${\cal K}^{\mu}$ by the relation
\begin{equation} 
\phi (x,{\cal A})-\tilde{\phi} (y,{\cal A})=-\int 
d^2z{\cal A}^{\mu}(z){\cal K}_{\mu}(z;x,y)\; ,
\label{defk} 
\end{equation}
and satisfying
\begin{equation} 
\partial^{\mu}_z{\cal K}_{\mu}(z;x,y)=e\left[\delta^{(2)}(x-
z)-\delta^{(2)}(y-z)\right]\; , 
\label{divk} 
\end{equation}
which is evident by virtue of the definitions~(\ref{gauge1},\ref{gauge2}) of  
$\phi$ and $\tilde{\phi}$. In this notation the operator~(\ref{oper}) is 
just: $\exp\left[\int d^2z{\cal K}^{\mu}(z;x,y)\delta/\delta J^{\mu}(z) 
\right]$ and simply shifts the argument of the $J$ dependent functional
\begin{equation}
\exp\left[\int d^2z{\cal K}^{\mu}(z;\cdot ,\cdot)\frac{\delta}{\delta 
J^{\mu}(z)}\right]F[J]=F[J+{\cal K}]\; .
\label{shift}
\end{equation}

Having found expression~(\ref{difx}) for the second derivative of 
$X$, and collecting~(\ref{zb}) together with~(\ref{prop}) we are in a 
position to write down the formula for the one-instanton contribution to 
the quark propagator
\begin{eqnarray}
&&S^{(1)}(x-y)=i\lambda e^{i\theta}\chi_0(x)\chi_0^+(y)\cdot\exp
\bigg\{\frac{i}{2}\int
d^2u\left[b^{(1)}(u)\Box\left(\Box+e^2/\pi\right)b^{(1)}(u)\right] 
\label{s1}\\
&&-\frac{i}{2}\int d^2ud^2w\left[{\cal K}^{\mu}(u;x,y) + 
\varepsilon^{\mu\alpha}\left(\Box+e^2/\pi\right)\partial_{\alpha} 
b^{(1)}(u)\right]\bigtriangleup_{\mu\nu}(u-w) \Big[{\cal 
K}^{\nu}(w;x,y) + \nonumber\\
&&\varepsilon^{\nu\beta}\left(\Box+e^2/\pi\right)\partial_{\beta} 
b^{(1)}(w)\Big]\bigg\}\; .\nonumber
\end{eqnarray}
In the exponent of~(\ref{s1}) all functions are perfectly known, so it is 
not difficult, although lengthy, to evaluate all expressions. We skip this 
calculation here and only note that all terms may be divided onto three 
groups: those not containing current ${\cal K}^{\mu}$,  linear, and 
finally quadratic in ${\cal K}^{\mu}$.
\begin{enumerate}
\item It is not difficult to check that both terms which do not depend on 
${\cal K}$ cancel each other.
\item The terms linear in ${\cal K}$ are strongly simplified if one notices 
that thanks to the presence of $\varepsilon^{\mu\nu}$ from the whole 
expression for $\bigtriangleup_{\mu\nu}$ only that proportional to metric 
tensor contributes. If one now exploit the known identity for gamma 
matrices in two dimensions 
$\gamma^{\beta}\gamma^5=\varepsilon^{\mu\beta}\gamma_{\mu}$ and 
observe that finally $\gamma^5$ may be replaced with -1, because of the 
structure of the matrical coefficient ($\chi_0\chi_0^+$)
in~(\ref{s1}), we see that the total effect of these terms is just to
cancel $\chi_0\chi_0^+$ (up to $\frac{1}{2\pi}$) together with the
coefficient $\lambda$. This was, naturally, expected since the conclusive 
formula should not contain any trace of the particular choice of $b^{(1)}$.
\item The calculation of the term quadratic in ${\cal K}$ is also 
elementary and the result turns out to be
\begin{equation}
\gamma_E + \frac{1}{2}\ln\frac{e}{2\sqrt{\pi}} + ie^2\beta (x-y)\; .
\label{I1}
\end{equation}
\end{enumerate}

This allows us to complete the formula~(\ref{s1}) for $S^{(1)}$. The 
twin calculation in the $k=-1$ instanton sector gives the analogous 
outcome with the reversed signs of parameter $\theta$ and matrix 
$\gamma^5$. Finally, the quark propagator with contributions from all 
sectors is
\begin{equation}
S(x)={\cal S}_0(x)e^{-ie^2\beta(x)}+ 
\frac{ie}{4\pi^{3/2}}\left(\cos\theta-i\gamma^5\sin\theta\right)e^{\gamma_E 
+ie^2\beta(x)}\; .
\label{s01}
\end{equation}

This proves that the second and the third term in~(\ref{fulls}), admitted by 
the Dyson-Schwinger equation, do actually materialize as a result of 
instanton effects. The last term does not reappear in~(\ref{s01}), and in 
fact may not do, by virtue of simple dimensional consideration. 
From~(\ref{fulls}) it is obvious that the constant $D_0$ has to be 
dimensionless and simultaneously it must tend to zero when 
$e^2\rightarrow 0$ (the free case). Naturally no such constant exists
in a world in which 
the only dimensionful constant is just $e^2$. From similar
dimensional analysis it is obvious that constants $B_0$ and $C_0$
in~(\ref{fulls}) must be linear in $e$ which is in fact the
case in~(\ref{s01}). One should note in this point that instanton
effects usully manifest themselves in a nonperturbative way, for
instance through the appearance of negative powers of the coupling
constant. As an example we can quote the one-instanton tunelling
amplitude in four-dimensional YM: $<0|1>\sim\exp(-8\pi^2/g^2)$~\cite{ynd}. In
the present case, however, we cannot construct $e^2$-dependent
dimensionless constants, and the intuition concerning nonperturbative
contributions should be confronted with dimensional considerations.
In the last section we come 
back to the formula~(\ref{s01}) and analyze it from the point of view of 
the quark condensate.

It should be also observed that other functions with two quark `legs', as for 
instance the vertex function, are still completely expressible through the 
propagator $S$ as discussed in~\cite{trjmn} and the relation
\begin{equation}
\left[S\Gamma_{\mu}S\right]^{(k)}(x,y;z)= iS^{(k)}(x-
y)\gamma^{\mu}\not\! \partial_z\bigtriangleup(y-z) -i\not\! 
\partial_z\gamma^{\mu}\bigtriangleup(x-z)S^{(k)}(x-y)\; ,
\label{vertk}
\end{equation}
as well as its counterparts for higher functions, hold separately in each 
instanton sector.

\section{Instanton contributions to the 4-fermion function}
\label{sec:4p}

In compliance with our former discussion, the expectation value of the 
product $\Psi\Psi\overline{\Psi}\overline{\Psi}$ acquires contributions 
from up to $k=\pm 2$ instanton sectors. As in the case of the propagator, 
the 4-fermion Green function in a certain specific sector is defined
by the functional derivative of the generating functional
\begin{equation}
G^{(k)}_{ab,cd}(x_1,x_2;x_3,x_4)=e^{ik\theta} \frac{1}{Z[0,0,0]} 
\left. \frac{\delta^4Z^{(k)}[\eta,\overline{\eta},J]} 
{\delta\overline{\eta}_a(x_1)\delta\overline{\eta}_b(x_2) 
\delta\eta_c(x_3) \delta\eta_d(x_4)}\right|_{\eta,\overline{\eta},J=0}\; .
\label{def4}
\end{equation}
For the sector $k=0$ the appropriate expression for the Green function 
was found in our previous work~\cite{trjmn}. We, therefore, begin with 
considering the case $k=1$ recalling here only the final result for $G^{(0)}$
\begin{eqnarray}
&&G^{(0)}_{ab;cd}(x_1,x_2;x_3,x_4)=\frac{1}{2}\bigg[{\cal 
S}_{0ac}(x_1-x_3){\cal S}_{0bd}(x_2-x_4)
+\left({\cal S}_0(x_1-x_3)\gamma^5\right)_{ac}\left({\cal S}_0(x_2-
x_4)\gamma^5\right)_{bd}
\bigg]\nonumber\\
&&\exp\bigg[ie^2(\beta(x_1-x_2) - \beta(x_1-x_3) - \beta(x_2-x_4)
-\beta(x_1-x_4)-\beta(x_2-x_3)+\beta(x_3-x_4))\bigg]\nonumber\\
&&+\frac{1}{2}\bigg[{\cal S}_{0ac}(x_1-x_3){\cal S}_{0bd}(x_2-
x_4)- \left({\cal S}_0(x_1-x_3)\gamma^5
\right)_{ac}\left({\cal S}_0(x_2-
x_4)\gamma^5\right)_{bd}\bigg]\nonumber\\
&&\exp\bigg[-ie^2(\beta(x_1-x_2) - \beta(x_1-x_3) - \beta(x_2-x_4) - 
\beta(x_1-x_4)-\beta(x_2-x_3)
+\beta(x_3-x_4))\bigg]\nonumber\\
&&-(c, x_3 \longleftrightarrow d, x_4)\; .\label{fin4f}
\end{eqnarray}
The evaluation of $Z^{(1)}$ has already been done in the previous 
section so we are able to immediately write down the following equation 
for $G^{(1)}$
\begin{eqnarray}
&&G^{(1)}_{ab,cd}(x_1,x_2;x_3,x_4)= i\lambda e^{i\theta} 
\Biggr\{\Biggr[\left(e^{-i\phi(x_1,\delta/i\delta J)}\chi_0(x_1)\chi_0^+(x_3) 
e^{i\tilde{\phi}(x_3,\delta/i\delta J)}\right)_{ac}
\Big(e^{-i\phi(x_2,\delta/i\delta J)} \nonumber \\
&&e^{-i\phi(x_2,\varepsilon^{\mu\nu}\partial_{\nu} b^{(1)})}{\cal
S}_0(x_2-x_4) 
e^{i\tilde{\phi}(x_4,\varepsilon^{\mu\nu}\partial_{\nu} b^{(1)})} 
e^{i\tilde{\phi}(x_4,\delta/i\delta J)}\Big)_{bd}- (c,
x_3\leftrightarrow d, x_4)\Biggr] 
-(a, x_1\leftrightarrow b, x_2)\Biggr\}\nonumber\\
&&\exp\biggr[\frac{i}{2}\int d^2xb^{(1)}\Box 
\left(\Box+e^2/\pi\right)b^{(1)}- \frac{i}{2}\int 
d^2xd^2y\left(J^{\mu}(x) + 
\varepsilon^{\mu\alpha}\left(\Box+e^2/\pi\right)\partial_{\alpha} 
b^{(1)}(x)\right)\nonumber\\
&&\bigtriangleup_{\mu\nu}(x-y) \left(J^{\nu}(y) + 
\varepsilon^{\nu\beta}\left(\Box+e^2/\pi\right)\partial_{\beta} 
b^{(1)}(y)\right)\biggr]\; ,\label{g1}
\end{eqnarray}
taken at $J=0$. We now recall the definition~(\ref{defk}) of the
current ${\cal K}$ 
from the preceding section and ${\cal J}$ introduced 
in~\cite{trjmn} by the relation
\begin{equation} 
\tilde{\phi} (x,{\cal A})-\tilde{\phi} (y,{\cal A})=-\int 
d^2z{\cal A}^{\mu}(z){\cal J}_{\mu}(z;x,y)\; . 
\label{defj} 
\end{equation}
The both currents satisfy the equation
\begin{equation}
{\cal K}^{\mu}(z;x,y)={\cal J}^{\mu}(z;x,y)-
2e\gamma^5\varepsilon^{\mu\nu}\partial^z_{\nu}\bigtriangleup(x-z)\; .
\label{kj}
\end{equation}
Using this notation we can rewrite~(\ref{g1}) in the form
\begin{eqnarray}
&&G^{(1)}(x_1,x_2;x_3,x_4)= i\lambda e^{i\theta} 
\left(\chi_0(x_1)\chi_0^+(x_3) \otimes {\cal S}_0(x_2-
x_4)\right)
\Bigg(\exp\left[\int d^2z{\cal K}^{\mu}(z;x_1,x_3)\frac{\delta}{\delta 
J^{\mu}(z)}\right]\nonumber\\
&&\otimes\exp\left[i\int d^2z {\cal J}_{\mu}(z;x_2,x_4) 
\varepsilon^{\mu\nu}\partial_{\nu}b^{(1)}(z)
+ \int d^2z{\cal J}^{\mu}(z;x_2,x_4)  
\frac{\delta}{\delta J^{\mu}}\right]\Bigg)\times\nonumber\\
&&\exp\bigg[\frac{i}{2}\int d^2xb^{(1)}\Box 
\left(\Box+e^2/\pi\right)b^{(1)}-\frac{i}{2}\int d^2xd^2y\left(J^{\mu}(x) + 
\varepsilon^{\mu\alpha}\left(\Box+e^2/\pi\right)\partial_{\alpha} 
b^{(1)}(x)\right)\times\nonumber\\
&& \bigtriangleup_{\mu\nu}(x-y)\big(J^{\nu}(y)+\varepsilon^{\nu\beta}
\left(\Box+e^2/\pi\right)\partial_{\beta} 
b^{(1)}(y)\bigg)\bigg] +\; {\rm antisymmetrization}\; ,\label{ga}
\end{eqnarray}
where the `antisymmetrization' is defined by substitutions
in~(\ref{g1}). The obvious 
matrical notation has been introduced above to simplify the expression 
and avoid explicitly writing four spinor indices. (One should keep in mind 
that both currents ${\cal K}$ and ${\cal J}$ are matrices in spinor space.)
The functional derivatives over current $J$, similarly as in the expressions 
of  the previous section, lead to shifts of $J$ and we obtain
\begin{eqnarray}
&&G^{(1)}(x_1,x_2;x_3,x_4)= i\lambda e^{i\theta} 
\bigg[\chi_0(x_1)\chi_0^+(x_3) \otimes {\cal S}_0(x_2-x_4) 
\exp\bigg(i\int d^2z {\cal J}_{\mu}(z;x_2,x_4)\times\label{gb}\\
&&\varepsilon^{\mu\nu}
\partial_{\nu}b^{(1)}(z)\bigg)\bigg]\exp\biggr[\frac{i}{2}\int
d^2xb^{(1)}\Box
\left(\Box+e^2/\pi\right)b^{(1)}- \frac{i}{2}\int 
d^2xd^2y\Big({\cal K}^{\mu}(x;x_1,x_3)\otimes\openone + \nonumber\\
&&\openone\otimes {\cal J}(x;x_2,x_4)+ 
\varepsilon^{\mu\alpha}\left(\Box+e^2/\pi\right)\partial_{\alpha} 
b^{(1)}(x)\Big)\bigtriangleup_{\mu\nu}(x-y) \Big({\cal 
K}^{\nu}(y;x_1,x_3)\otimes\openone + \nonumber\\
&&\openone\otimes {\cal J}(y;x_2,x_4)+ 
\varepsilon^{\nu\beta}\left(\Box+e^2/\pi\right)\partial_{\beta} 
b^{(1)}(y)\Big)\biggr]+\; {\rm antisymmetrization}\; ,\nonumber
\end{eqnarray}
In the last exponent of~(\ref{gb}) we recognize the expressions similar to 
those of~(\ref{s1}) although the tensor structure is now much more 
perplexed. Nevertheless all functions are known and the evaluation 
of~(\ref{gb}) is only a matter of patience. We do not intend to go into 
details and only indicate the main points.
\begin{enumerate}
\item The term quadratic in ${\cal K}$ is just that of the previous section 
multiplied (tensor product) by $\openone$.
\item The term quadratic in ${\cal J}$ was calculated in our previous 
work~\cite{trjmn}.
\item  The two terms containing squares of $b^{(1)}$ cancel each other.
\item The terms containing products of ${\cal J}$ and $b^{(1)}$ cancel 
with the first exponent in~(\ref{gb}).
\item The terms containing products of ${\cal K}$ and $b^{(1)}$ cancel 
with the appropriate part of the coefficient $\chi_0(x)\chi_0^+(y)$ (see 
Section~\ref{sec:2pinst}).
\item The product terms of  both currents ${\cal K}$ and ${\cal J}$ may 
be evaluated in the straightforward way and we obtain
\begin{equation}
ie^2\gamma^5\otimes\gamma^5\left[\beta(x_1-x_4) - \beta(x_1-x_2) - 
\beta(x_2-x_3) + \beta(x_3-x_4)\right]\; .
\label{jk}
\end{equation}
\end{enumerate}
With all above taken into account we can now the arising exponent 
in~(\ref{gb}) give the following simple form
\begin{eqnarray}
\exp\left[a\openone\otimes\openone+b\gamma^5\otimes\gamma^5\right] = &&
\frac{1}{2}\left(\openone\otimes\openone - 
\gamma^5\otimes\gamma^5\right)\exp(a-b) +\nonumber\\ 
&& \frac{1}{2}\left(\openone\otimes\openone + 
\gamma^5\otimes\gamma^5\right)\exp(a+b)\; ,
\label{simp}
\end{eqnarray}
where $a$ and $b$ are certain functions expressible by combinations of 
$\beta$'s.

The calculations for $k=-1$, which are practically identical, allow us to 
write down the final expression for the one instanton contributions to $G$
\begin{eqnarray}
&&G^{(+1)}(x_1,x_2;x_3,x_4) + G^{(-1)}(x_1,x_2;x_3,x_4) = -
\frac{ie}{8\pi^{3/2}}e^{\gamma_E}\left(\cos\theta-
i\gamma^5\sin\theta\right)\otimes{\cal S}_0(x_2-x_4)\times\nonumber\\
&&\biggr[\left(\openone\otimes\openone - 
\gamma^5\otimes\gamma^5\right) e^{ie^2\left[\beta(x_1-x_3) - 
\beta(x_2-x_4) - \beta(x_1-x_4) + \beta(x_1-x_2) + \beta(x_2-x_3) - 
\beta(x_3-x_4)\right]} + \label{gp1m1}\\
&&\left(\openone\otimes\openone + 
\gamma^5\otimes\gamma^5\right) e^{ie^2\left[\beta(x_1-x_3) - 
\beta(x_2-x_4) + \beta(x_1-x_4) - \beta(x_1-x_2) - \beta(x_2-x_3) + 
\beta(x_3-x_4)\right]}\biggr]\nonumber\\
&& +\; {\rm antisymmetrization}\; . \nonumber
\end{eqnarray}

In the sector $k=2$ we use again the formula~(\ref{def4}). There 
are now two zero modes. This means that after functional integration over 
Grassmann variables, analogous to that performed to obtain~(\ref{x1}), 
the product of two $\eta$'s and two $\overline{\eta}$'s appears. These 
four sources have to saturate all four derivatives in~(\ref{def4}), since 
otherwise we would get zero after setting $\eta,\overline{\eta},J=0$. That 
in turn means that, in the formula for $G^{(2)}$, the current ${\cal J}$ 
will not appear and
\begin{eqnarray}
&&G^{(2)}(x_1,x_2;x_3,x_4)= -\lambda^2e^{2i\theta} \chi_0(x_1)\chi_0^+(x_3) 
\otimes \chi_1(x_2)\chi_1^+(x_4)\exp\biggr[\frac{i}{2}\int d^2xb^{(2)}\Box 
\left(\Box+e^2/\pi\right)b^{(2)} \nonumber\\
&&- \frac{i}{2}\int 
d^2xd^2y\Big({\cal K}^{\mu}(x;x_1,x_3)\otimes\openone 
+\openone\otimes {\cal K}(x;x_2,x_4)+ 
\varepsilon^{\mu\alpha}\left(\Box+e^2/\pi\right)\partial_{\alpha} 
b^{(2)}(x)\Big)\nonumber\\
&&\bigtriangleup_{\mu\nu}(x-y) \Big({\cal 
K}^{\nu}(y;x_1,x_3)\otimes\openone + \openone\otimes {\cal 
K}(y;x_2,x_4)+ 
\varepsilon^{\nu\beta}\left(\Box+e^2/\pi\right)\partial_{\beta} 
b^{(2)}(y)\Big)\biggr]\label{g2b}\\
&& +\; {\rm antisymmetrization}\; .\nonumber
\end{eqnarray}
We are not going to repeat below the steps leading to the final result, 
since the modification of previous formulae is only slight. The 2-instanton 
configurations contribute then in the following way
\begin{eqnarray}
&&G^{(+2)}(x_1,x_2;x_3,x_4) + G^{(-2)}(x_1,x_2;x_3,x_4) = -
\frac{e^4}{256\pi^4}e^{4\gamma_E}\bigg[e^{2i\theta}\left(\openone - 
\gamma^5\right)\otimes\left(\openone - \gamma^5\right)(x_2^0+x_2^1) 
\nonumber\\
&&(-x_4^0+x_4^1) + e^{-2i\theta}\left(\openone + 
\gamma^5\right)\otimes\left(\openone + \gamma^5\right)(-x_2^0+x_2^1) 
(x_4^0+x_4^1)\bigg]\exp\Big\{ie^2\big[\beta(x_1-x_4) +\label{gp2m2}\\ 
&&\beta(x_2-x_3) +\beta(x_1-x_2) + 
\beta(x_3-x_4) + \beta(x_1-x_3) + \beta(x_2-x_4)\big]\Big\} + \; {\rm 
antisymmetrization}\; . \nonumber
\end{eqnarray}
If we now expand the tensor notation of~(\ref{gp1m1}) and~(\ref{gp2m2}) 
into explicit spinor indices and perform the full antisymmetrization (which 
also restores the apparently broken Lorentz invariance) as defined 
by~(\ref{g1}) we can collect together all the contributions
\begin{eqnarray}
G_{ab,cd}(x_1,x_2;x_3,x_4) =&& G^{(0)}_{ab,cd}(x_1,x_2;x_3,x_4) + 
G^{(+1)}_{ab,cd}(x_1,x_2;x_3,x_4) + G^{(-
1)}_{ab,cd}(x_1,x_2;x_3,x_4) +\nonumber\\ 
&& G^{(+2)}_{ab,cd}(x_1,x_2;x_3,x_4) +G^{(-2)}_{ab,cd}(x_1,x_2;x_3,x_4)\; .
\label{gfull}
\end{eqnarray}

Since we now dispose the complete formula for the 4-fermion function, 
we can extend the analysis carried out in~\cite{trjmn}, regarding the 
possible existence of the Schwinger pole in the $t$-channel (the
quark-antiquark scattering), over 
contributions from higher topological sectors. To this goal we identify the 
appropriate co-ordinates of incoming and outgoing particles:
$u\stackrel{\rm def}{=} 
x_1=x_3$, $v\stackrel{\rm def}{=} x_2=x_4$ and perform the Fourier
transform of the expression in the new variable $z\stackrel{\rm
def}{=} v-u$. This identification (which should be treated as an
appropriate limit) has to be done in a symmetrical way as 
described in the quoted paper. After these manipulations we see that the 
formula~(45) from~\cite{trjmn} acquires additional terms from the $k\neq 
0$ instanton sectors
\begin{eqnarray}
G_{polar}(k)=&&-\frac{i}{4\pi}\gamma^0\otimes\gamma^0 
\frac{(k^1)^2}{(k^2-e^2/\pi)}+\frac{ie}{4\pi^{3/2}}e^{\gamma_E} \left( 
\gamma^0\otimes\gamma^5e^{-i\theta\gamma^5} -
\gamma^5e^{-i\theta\gamma^5}\otimes\gamma^0\right)\times\nonumber\\ 
&&\frac{k^1}{(k^2-
e^2/\pi)}+ \frac{ie^2}{8\pi^2}e^{2\gamma_E}e^{-i\theta\gamma^5}\otimes  
e^{-i\theta\gamma^5}\cdot \left[\openone\otimes\openone 
+\gamma^5\otimes\gamma^5\right]\; .
\label{pol012}
\end{eqnarray}

We would like to point out here certain observation. If one 
considers the time ordered product of two $\Psi$'s and two 
$\overline{\Psi}$'s then, for some choice of time arguments, one can 
introduce a complete set of (out) states obtaining for instance 
$$
\sum_n<0|\Psi(x_1)\overline{\Psi}(x_3)|n><n|\Psi(x_2) 
\overline{\Psi}(x_4)|0>\; .
$$
The set $|n>$ is here a Fock set of massive Schwinger
bosons since these are the only asymptotic states. 
The 1-boson contribution, which leads to the appearance of a pole at 
$k^2=e^2/\pi$ is expressible, thanks to the formulae of the LSZ 
formalism, through the vertex function. This in turn means, by virtue
of the 
equation~(\ref{vertk}), that each of the amplitudes above have the 
contributions from both $k=\pm 1$ sectors. In consequence, the polar part 
of $G$ should, beside sectors $+1+1$ and $-1-1$, contain also traces of  
sectors $+1-1$ and $-1+1$. While calculated explicitly this leads to the 
substitution $\gamma^5\otimes\gamma^5$ for 
$\openone\otimes\openone$ in the last term of~(\ref{pol012}). Why 
such contributions appear neither in $G^{(0)}$ nor in $G^{(2)}$ remains 
for us presently unclear.

\section{Conclusions}
\label{sec:conclus}

The expressions for the Green functions obtained in sections~\ref{sec:2p} 
and~\ref{sec:4p} show that the Schwinger Model may be explicitly solved also 
with all topological effects taken into account. As argued in~\cite{trjmn},
all higher functions (with no more than 4 external fermion `legs'), thanks 
to the two Ward identities, which are preserved in the nonzero instanton 
sectors too, are the combinations of those found above. 

From the formulae~(\ref{s01}) and~(\ref{gfull}) together
with~(\ref{fin4f},\ref{gp1m1},\ref{gp2m2}) we can obtain, by taking the
appropriate limits, the values of the condensates and verify
the cluster property. We do not treat, in this work, the restoration
of the cluster property as a support for the use of the
$\theta$-vacuum instead of a topological one since this fact is well
known and needs no verification. We rather show that the obtained
formulae are selfconsistent and agree (in this special limit) with
former results of other authors.

The values of condensates 
\begin{equation}
{\cal V}\stackrel{\rm def}{=}<\overline{\Psi}(x)\Psi(x)>_{vac}\;
,\;\;\;\;\; {\cal 
V}_A\stackrel{\rm def}{=}<\overline{\Psi}(x)\gamma^5\Psi(x)>_{vac}\; ,
\label{convdef}
\end{equation}
may be easily obtained from eq.~(\ref{s01}). If we recall that 
$\beta(0)=0$ we get
\begin{equation}
{\cal V}=\frac{e}{2\pi^{3/2}}e^{\gamma_E}\cos\theta\; ,\;\;\;\;\; {\cal 
V}_A=-\frac{ie}{2\pi^{3/2}}e^{\gamma_E}\sin\theta \; ,
\label{condval}
\end{equation}
which agrees with the results of other authors~\cite{cadam1,gmc} and 
constitutes the 1-instanton contribution. To verify the
clusterization property let us now consider the objects 
defined by
\begin{equation}
{\cal W}(x,y)\stackrel{\rm def}{=}<\overline{\Psi}(x)\Psi(x) 
\overline{\Psi}(y)\Psi(y)>_{vac}\; ,\;\;\;\;\; {\cal 
W}_A(x,y)\stackrel{\rm def}{=}<\overline{\Psi}(x)\gamma^5\Psi(x) 
\overline{\Psi}(y)\gamma^5\Psi(y)>_{vac}\; ,
\label{conwdef}
\end{equation}
The $k=\pm 1$ does not contribute here since the trace of the product
of an odd number 
of gamma matrices is zero and that is what we obtain while exploiting
equation~(\ref{gp1m1}). From the sector $k=0$ we obtain
\begin{eqnarray}
{\cal W}^{(0)}(x,y)&=&\frac{e^2}{8\pi^3} 
e^{2\gamma_E+2K_0(\sqrt{-e^2(x-y)^2/\pi})}\; , \label{w0}\\
{\cal W}_A^{(0)}(x,y)&=&-\frac{e^2}{8\pi^3} 
e^{2\gamma_E+2K_0(\sqrt{-e^2(x-y)^2/\pi})}\; ,
\end{eqnarray}
and for $k=\pm 2$ the result is
\begin{equation}
{\cal W}^{(+2;-2)}(x,y)={\cal W}_A^{(+2;-2)}(x,y)= 
\frac{e^2}{8\pi^3} e^{2\gamma_E - 2K_0(\sqrt{-e^2(x-
y)^2/\pi})}\cos 2\theta\; .
\label{w2}
\end{equation}
In particular in the limit $(x-y)^2\rightarrow -\infty$, where 
$K_0\rightarrow 0$, we have
\begin{equation}
{\cal W}=\frac{e^2}{8\pi^3}e^{2\gamma_E}(1+\cos 2\theta)\; , \;\;\;\;\; 
{\cal W}_A=\frac{e^2}{8\pi^3}e^{2\gamma_E}(-1+\cos 2\theta)\; ,  
\label{wres}
\end{equation}
and thanks to the identity $\cos 2\theta=2\cos^2\theta -1= 1-2 
\sin^2\theta$ the clusterization reappears
\begin{equation}
{\cal W}={\cal V}\cdot{\cal V}\; , \;\;\;\;\; {\cal W}_A={\cal 
V}_A\cdot{\cal V}_A\; .
\label{clus}
\end{equation}

We would like to devote now a few words to the $\theta$ dependence
of the Green functions we obtained. It is well
known~\cite{huang,raja} that in the case of massless fermions, thanks
to the chiral invariance of the Lagrangian, the parameter $\theta$
may be gauged away. This happens because of the close connection
between the chiral anomaly and the winding number. That it can
actually be done is confirmed by our
formulae~(\ref{s01},\ref{fin4f},\ref{gp1m1},\ref{gp2m2}). For each
product of $\Psi$ and $\overline{\Psi}$ in a Green function there
appears a factor $e^{-i\theta\gamma^5}$ in the final formulae 
which might obviously be
cancelled by the appropriate chiral gauge transformation performed on
the fermion field.

\acknowledgements
The author would like to thank to Professors J. Namys{\l}owski and K. 
Meissner for interesting and inspiring discussions.


\begin{references}
\bibitem{com1} The source term should be written in that way 
that the functional derivative  $\frac{\delta}{i\delta J}$ act only on 
$e^{i\int d^2x J^{\mu}A^{(0)}_{\mu}}$.
\end{references}
\end{document}